\documentclass{elsart}

\usepackage{natbib}

\usepackage{amssymb}

\begin{document}
\input{epsf}
\renewcommand{\labelenumi}{\arabic{enumi}}

\begin{frontmatter}

% Title, authors and addresses

% use the thanksref command within \title, \author or \address for footnotes;
% use the corauthref command within \author for corresponding author footnotes;
% use the ead command for the email address,
% and the form \ead[url] for the home page:
% \title{Title\thanksref{label1}}
% \thanks[label1]{}
% \author{Name\corauthref{cor1}\thanksref{label2}}
% \ead{email address}
% \ead[url]{home page}
% \thanks[label2]{}
% \corauth[cor1]{}
% \address{Address\thanksref{label3}}
% \thanks[label3]{}

\title{GOTPM: A Parallel Hybrid Particle-Mesh Treecode}

% use optional labels to link authors explicitly to addresses:
% \author[label1,label2]{}
% \address[label1]{}
% \address[label2]{}

\author[astro,cita]{John Dubinski\thanksref{dub}},
\author[cita,seoul]{Juhan Kim\thanksref{kjhan}},
\author[seoul]{Changbom Park\thanksref{cbp}}, and
\author[cita]{Robin Humble\thanksref{rjh}}
\thanks[dub]{e-mail: dubinski@astro.utoronto.ca}
\thanks[kjhan]{e-mail: kjhan@astro.snu.ac.kr}
\thanks[cbp]{e-mail: cbp@astro.snu.ac.kr}
\thanks[rjh]{e-mail: rjh@cita.utoronto.ca}

\address[astro]{Department of Astronomy and Astrophysics, University of Toronto}
\address[cita]{Canadian Institute for Theoretical Astrophysics}
\address[seoul]{Astronomy Program, School of Earth and Environmental Sciences, 
Seoul National University, Korea}

\begin{abstract}
We describe a parallel, cosmological N-body code based
on a hybrid scheme using the particle-mesh (PM) 
and Barnes-Hut (BH) oct-tree algorithm.  We call the algorithm 
GOTPM for Grid-of-Oct-Trees-Particle-Mesh.   The
code is parallelized using the Message Passing Interface (MPI)
library and is optimized to run on Beowulf
clusters as well as symmetric multi-processors. 
The gravitational potential is determined on a mesh using a
standard PM method with particle forces determined through
interpolation.   The softened PM force is corrected for
short range interactions using a grid of localized BH trees throughout
the entire simulation volume in a completely analogous
way to P$^3$M methods.
This method makes no assumptions about the local density 
for short range force
corrections and so is consistent with the results of the P$^3$M
method in the limit that the treecode opening angle parameter,
$\theta \rightarrow 0$.

The PM method is parallelized using
one-dimensional slice domain decomposition.  Particles are
distributed in slices of equal width to allow mass assignment
onto mesh points.  The Fourier transforms in the PM method are
done in parallel using the MPI implementation of the FFTW
package. Parallelization for the tree force corrections is
achieved again using one-dimensional slices but the width of each
slice is allowed to vary according to the amount of computational
work required by the particles within each slice to achieve
load balance. The tree force corrections dominate the
computational load and so imbalances in the PM density assignment
step do not impact the overall load balance and performance
significantly.  The code performance scales well to 
128 processors and is significantly better than competing methods.
We present preliminary results from simulations run on different
platforms containing up to $N=1G$ particles to verify the code. 

%The overall performance of this code has been tested under two
%parallel environments: a Linux Beowulf cluster and a Compaq GS320
%SMP. The scaling of execution times with a various number of CPU's
%shows that this PM/TREE code has a good performance applicable to
%massive parallel machines having a few hundred CPU's. We have
%profiled  the PM part of the code with $N=1G$ particle simulation
%in a $\Lambda=0.7+\Omega=0.3$ CDM cosmological model in a $L=630$
%$h^{-1}$ Mpc volume. The full PM/tree hybrid code was tested using
%a $N=128M$ particle run in a smaller $L=65$ $h^{-1}$ Mpc volume
%from $z=71$ to $z=0$. We find that this code is an order of
%magnitude faster than a parallel, Ewald cosmological treecode and
%comparable in performance to similar parallel codes using the PM
%method plus some other method of short range force correction. The
%code performs well in the strongly clustered regime where $P^3M$
%and other methods face load balancing difficulties.  Since the
%tree-building process is localized to a few mesh distances
%multiple time-stepping is fairly easy to implement.

\end{abstract}

\begin{keyword}
% keywords here, in the form: keyword \sep keyword
methods: N-body simulations \sep methods: numerical \sep cosmology: dark matter

% PACS codes here, in the form: \PACS code \sep code

\end{keyword}

\end{frontmatter}

\section{Introduction}

As we enter the era of precision cosmology,
the need for higher resolution cosmological simulations has never been
clearer.
Recent determinations of the cosmological parameters to accuracies of a few
percent by the WMAP mission (Bennett et al 2003) will entirely transform 
the industry of cosmological N-body simulations.  
Since the initial conditions are now well-defined,
the focus will be on seeking detailed quantitative consistency with the 
prevailing cosmological
paradigm with observations of the large-scale structure through clustering
and lensing studies and detailed analysis of the structure and dynamics 
of galaxy clusters and galaxies themselves.  
During the last decade, there has been a huge growth in
computational power brought on by steady increase in processor
speed and the integration of processors into massively parallel
architectures. Numerical cosmologists and dynamicists have
developed parallel, N-body codes to take advantage these
architectures and there are now a variety of codes
that are being used to simulate the formation of large-scale
structure and galaxies.   In this paper, we describe 
a new parallel code 
that is adaptable and scalable to the next generation
of parallel supercomputers.

Two classes of cosmological N-body codes are now used widely and are based on 
either the Particle-Mesh (PM) algorithm (Hockney and Eastwood 1981)
or the Barnes-Hut Tree algorithm (1986).
PM codes are the fastest available taking advantage of the fast Fourier
transform algorithm (FFT) to solve Poisson's equation. 
While useful for some cosmological applications, pure PM codes fail to resolve structures on
scales smaller than the mesh size.  
Different approaches have been taken
increase the force resolution on the sub-mesh scale.  The simplest approach
is the P$^3$M algorithm where the mesh softened force is
corrected at short range using direct summation of pairwise forces 
between particles (e.g. Hockney and Eastwood 1981, Efstathiou et al
1985).   
As systems become clustered, the P$^3$M algorithm scales as O(N$_h^2$) 
where N$_h$ is the size of the largest halos
limiting its application to small N or very large cosmological volumes
where halos contain modest numbers of particles.
Another approach is to lay down refined sub-meshes on regions of high
density to compute force refinements.  This can be done hierarchically and
leads to accurate force determination and code speed up.
Couchman's (1991) AP$^3$M algorithm used in the Hydra code has been highly
successful.  A slight disadvantage of this approach is that a density
threshold criterion is necessary to lay down sub-meshes and there may be
discrete changes in force resolution across arbitrarily set boundaries that
have unpredictable effects.
An alternative mesh-refinement scheme is provided by 
Kravtsov et al. (1997) in
the ART algorithm that lays down refined meshes as particles begin
to crowd into mesh cells.
A completely different approach uses the BH tree algorithm modified to
include periodic boundary conditions using 
Ewald's method (Bouchet, Hernquist 
and Suto 1991).  This method has the advantage that forces are accurately
determined at all stages of the calculation on all scales.  The penalty is
that tree codes generally have much larger memory requirements and require
significantly more computation to achieve a similar force accuracy than PM
based algorithms.  Despite these problems, tree codes based on Ewald's method
have been easier to parallelize then AP$^3$M
are currently very competitive in many applications 
(e.g. Dav\'e et al 199?, Springel et al. 2001).

A natural evolution of cosmological codes is the developments of hybrids
that can incorporate the high efficiency of PM codes for long range force
determination and the advantage of tree codes for short range force
corrections.  Xu (1995) developed a hybrid Tree/PM (TPM) algorithm that 
is in the spirit of AP$^3$M and the algorithm has since been enhanced by
Bode et al. (2000).  BH trees are used in place of refining 
sub-meshes and
constructed around dense clustered regions to determine short range
force corrections.  
This method differs from the Ewald-method treecodes 
in that short range forces drop to zero at a few mesh lengths so trees
only need to be built locally to calculate the force corrections.
The algorithm has also been parallelized and is scalable to large processor
numbers (e.g. Bode and Ostriker 2003).   Bagla has also introduced a
hybrid TREEPM algorithm that breaks the forces up into short and a long
range parts using Ewald's original formalism.   The main difference from
TPM is that short range forces are corrected using the BH tree method
throughout the simulation volume with no need to specify a density 
threshold for laying down a tree.

In this paper, we describe a new Tree/PM hybrid that is in the same
spirit as Bagla's method but with parallelization.  We call this new algorithm
GOTPM standing for Grid-of-Oct-Trees-Particle-Mesh. 
In \S 2, we describe the details of the algorithm including a new parallel 
PM method based on the FFTW fast Fourier transform package (Frigo et al. 1998)
and a method for
computing the force refinements using localized BH trees.  Since tree
refinements are confined to a few mesh lengths, the process of building
trees within the simulation volume can be done sequentially leading to
large savings in memory compared with Ewald tree codes.  We also discuss
how to load balance the short range force correction calculations. 
In \S 3, we
present measures of the force accuracy and timing tests on a variety of
machines to measure the code's performance and parallel scalability.    
In \S 4, we describe some
preliminary results of
various simulations with the largest
containing N=$1024^3$ particles. 
In \S 5, we discuss the future incorporation of multi-timestepping 
and smoothed particle hydrodynamics  and algorithmic improvements
that will permit simulations containing $N=8-64G$ particles on next generation
parallel machines.

\section{Parallel TREEPM}

\subsection{Parallel PM}

The PM algorithm is well documented so we highlight here the features
of our implementation and method of parallelization.
We use the PM code developed by Park (1990, 1997) 
as our starting point
and it follows the usual steps:
\begin{enumerate}

\item We determine a density field on
a mesh from the particle
distribution using the triangular-shaped cloud (TSC) mass assignment scheme.
\item A Fast Fourier transform (FFT) of the density field is calculated and
then multiplied by the appropriate Green's function for the TSC scheme to
determine the transform of the potential field.
\item An inverse FFT determines the potential field in real space.
\item We interpolate the potential field to determine the
gravitational force at each particle position using a 4-point finite
difference approximation (FDA).
\end{enumerate}

We have parallelized this algorithm using a simple, slab domain
decomposition of the simulation cube using the Message-Passing Interface
(MPI) library.  Particles are first distributed in
slabs of equal width with an integral number of mesh divisions.
We determine the density field using the TSC method in each slab in each
processor.
Since the TSC method weights the particle mass to each of the 
surrounding 27 grid points
from the nearest grid point, we need to communicate the contribution to
the density field from particles near the boundaries of the slabs.
We achieve this by binning the mass on ghost slices of mesh that are
communicated to their neighbours and summed on to the corresponding local
slices to find the correct density
field quantities.

Once the density fields on slabs in each processor are determined we then 
use the FFTW library 
(Frigo et al. 1998)
with MPI extensions to calculate the FFTs used 
to obtain the gravitational potential.
The FFTW package requires that the complete mesh be divided in slabs 
of equal width so we can determine the gravitational potential immediately
with a parallel FFT.  With the potential, we then calculate the particle
forces in each slab using FDA interpolation.
Again, we need to import images of mesh slices at the boundaries to
determine forces on particles.  We communicate the 3 or 4 mesh slices 
from the front and back sides of the slab and then calculate the force on 
each particle in each processor in parallel.   Our implementation assumes
all neighbouring mesh slices are in adjacent slabs so the slab width can
be no smaller than 4 mesh divisions.  This sets a limits on the number of
processors that can be used in a parallel run e.g. only 128 processors can
be used with a $512^3$ mesh.  Although the code will work with thin slabs,
we find in practice below that problems can arise with memory imbalance in
the tree algorithm if slices are too thin so slabs with more
than 8 mesh divisions are preferred.

\subsection{Force Shaping}

\begin{figure}
\epsfxsize\hsize\epsffile{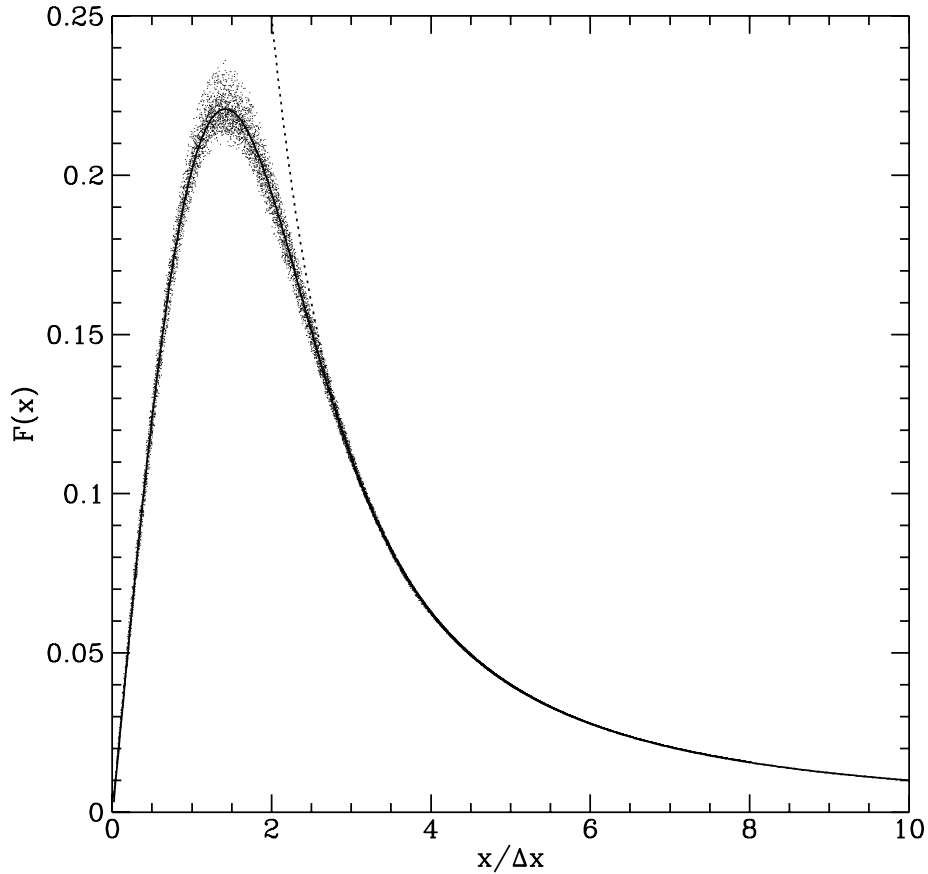}
\caption{
The PM force is calculated in a $N_{mesh}=256$ box at arbitrary positions around
a randomly positioned test particle in a homogeneous background
and plotted revealing the scatter due to force anisotropies imposed
by the mesh.  The solid line is
the best fit to the PM force according to our fitting function.
The force is normalized for unit mass and plotted versus
the separation in mesh cell sizes.
The dashed line represents the ideal Newtonian force that
adaptive refinements try to achieve.
}
\label{fig-pmfshape}
\end{figure}

\begin{figure}
\epsfxsize\hsize\epsffile{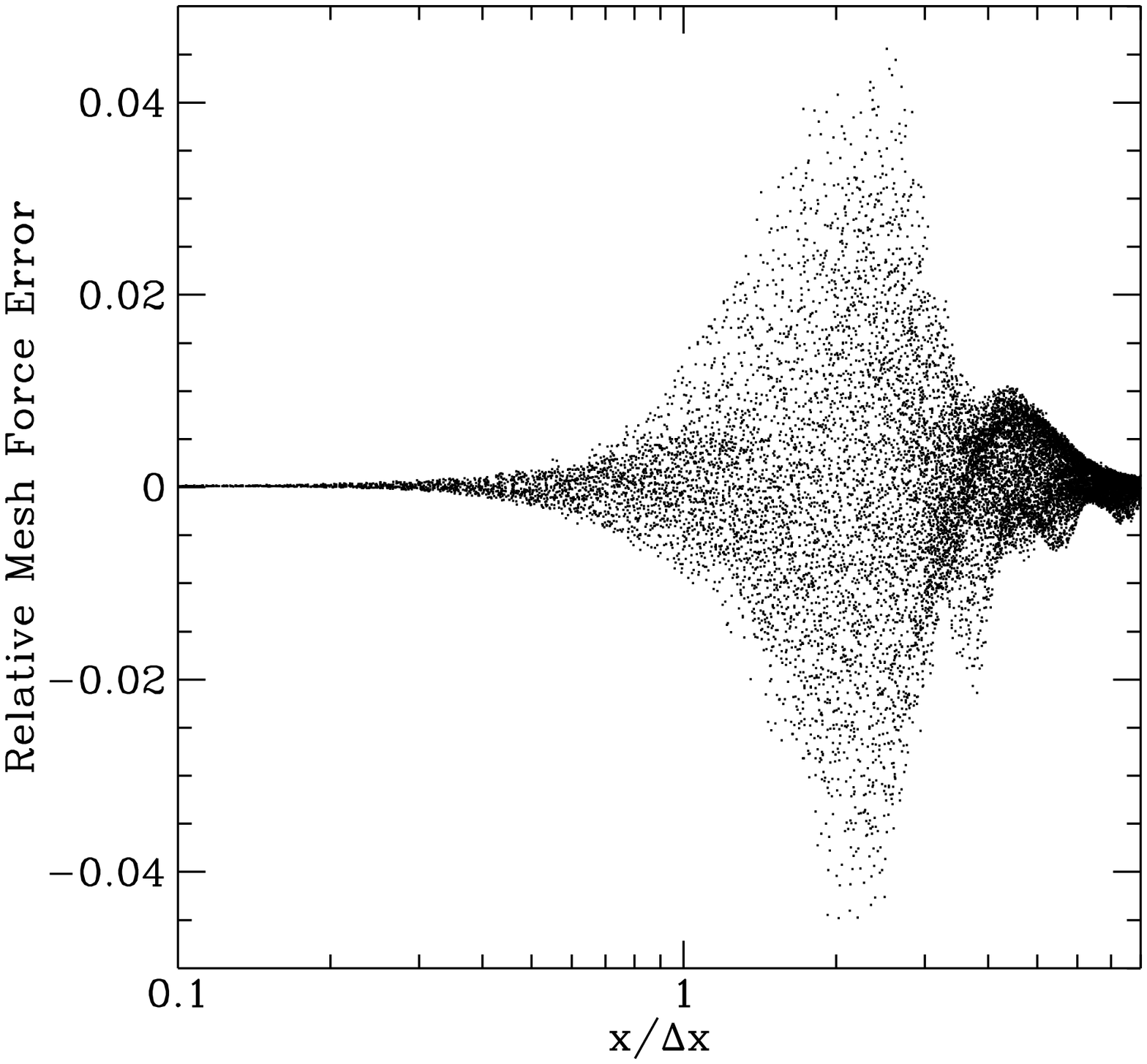}
\caption{
Scatter in the relative error in the mesh force + short range
correction force versus separation in mesh units
at randomly distributed points from a randomly
distributed point mass.  Anisotropies in the mesh force lead
to a large scatter at the mesh scale.
}
\label{fig-rerror}
\end{figure}

In the complete algorithm, we treat the PM force as the long range
contribution to the total force and determine the short range contribution 
using the tree algorithm.  The PM method softens the Newtonian potential
within a few mesh divisions and also introduces some anisotropy to the
force on the mesh size scale.  We follow the lead of other codes and
smooth the PM potential
using a Gaussian filter with
an e-folding radius of 0.9 times
the mesh size.  To characterize the resultant PM force, we calculate
the force many times at a point from a single particle placed 
at random positions within the mesh in a otherwise homogeneous density field.  
Figure \ref{fig-pmfshape} shows usual
behaviour with an asymptotic Newtonian force with significant scatter due 
to anisotropy at the maximum around 
$r=1.4 \Delta x$ where $\Delta x$ is the mesh cell size.
Our strategy is to fit these points with a smooth function of radius which
asymptotically behaves as $1/r^2$ with a best fit shown in Figure \ref{fig-pmfshape}.
After some experimentation, we settled on the rather complex fitting
function,
\begin{eqnarray}
\begin{array}{ll}
F(x) = &a\tanh(bx)/x^2 - ab /(x\cosh^2(bx)) \\
        & +cx(1+dx^2)\exp(dx^2) \\
        & +e(1+fx^2+gx^4)\exp(hx^2),
\end{array}
\end{eqnarray}
and find it successfully fits the data.
The short range force is simply determined as the
difference between this smooth empirical function and the Newtonian
gravitational force.  We also determined the scatter in the expected
relative force error by laying down a single particle randomly within
the grid and measuring the acceleration due to the mesh plus correction
at random points.  Figure \ref{fig-rerror} show that the scatter peaks
at a separation of about 2 mesh cell sizes revealing the anisotropy introduced by
working on a regular cubic mesh.  These errors are intrinsic to PM methods
and we see below in more general situations they introduce a tail in the
acceleration error distribution in the 1-2\% range.

\subsection{Correcting Forces with BH Trees}

The parallel PM force is very efficient at determining long range 
gravitational forces but because of the finite mesh size the forces
from nearby particles are highly softened.
The standard way to overcome this deficiency is to
define a new correction force law for local particles that is the difference
between the desired Newtonian force and the force law determined by the PM
method, $F_{c} = F_{N} - F_{PM}$.  The correction force has the property that it is nearly
Newtonian on the sub-mesh scales but then falls to zero beyond 
four mesh cell sizes where the PM forces become nearly Newtonian.
The gravitational potential is also softened using a Plummer law on the subgrid 
scale.  We typically use a softening radius of $5\times 10^{-5}L$, where L is
the boxsize i.e. so typical simulations are equivalent in resolution 
to a PM method with a $20000^3$ mesh.

Three approaches have been used to determine $F_c$ in PM methods.  Direct force evaluation
using the particle-particle (P$^2$) method (Hockney and Eastwood 1981), 
adaptive particle sub-meshes with P$^2$(Couchman 1991), and tree methods (Xu 1995, Bode et al. 2000,
and Bagla 2000).
Direct force evaluation is the most general method and is easily
parallelizable.  Unfortunately, when clustering is strong
the method grinds to
a halt because of the $O(N^2)$ nature of the method.
Couchman's (AP$^3$M) method relies on laying a local refined mesh
on regions where strong clustering is occuring and calculating correction
forces from this sub-mesh.  It is highly efficient in serial codes but
has proved difficult to parallelize.  
Finally,
Xu's hybrid TPM algorithm works in a similar way to AP$^3$M 
but a Barnes-Hut tree is built locally on clustered regions.  
This method has been parallelized and performs well (e.g. Bode et al. 2003).
A disadvantage of both these methods is that 
one requires an arbitrary
criterion to establish whether a mesh volume contains clustered material
that should be refined. Furthermore,
boundaries between finely resolved and more
coarsely resolved volumes are sharp and it is not clear if there are 
physical consequences to assigning these boundaries.

Bagla (2000) provides a way to characterize both the short and long range forces
in his TREEPM algorithim and is completely analogous to P$^3$M in that all local forces 
everywhere within the volume are corrected.  We describe here a parallel version of this method
that builds on the parallel PM algorithm above and incorporates most
of the efficient tree-building and tree walking software 
of Dubinski's (1996) parallel tree code PARTREE.
%Usually, use a critical density contrast (?) - uncertainty about
%continuity of forces acting on material falling into a cluster - also
%resolution of fine substructure is in doubt - what about galaxies forming
%in voids(?)
We call our new parallel algorithm GOTPM for Grid of-Oct-Trees-Particle-Mesh.  
This new code has the generality of a $P^3M$ code
in that all forces throughout the simulation volume are calculated accurately on the submesh 
scale with
no density criteria imposed for the addition of refinement.
The equivalent of mesh refinement is done automatically by building trees everywhere
within the simulation volume at every step.  We show below that this algorithm also
is easy to load balance using algorithms taken from pure parallel tree codes.

\subsection{GOTPM: A New Tree/PM Hybrid Algorithm}

It is easiest to describe this algorithm by comparing it directly to
the P$^3$M method.  The $P^2$ correction force on a
particle is determined by summing directly over particles within a sphere
centred on the particle which extends out to 2-3 mesh cell sizes. 
Beyond this boundary corrections on the mesh force are 
small, generally less than
1\%,  and are so neglected.  
The problem with $P^2$ correction, however, is
that the algorithm scales inherently as $O(N_h^2)$ and a code is therefore slowed down
considerably when clustering occurs.  In constrast, the tree algorithm is $O(N\log N)$
and generally requires about 2 to 3 times the amount of work when going from a 
homogeneous to a clustered state.

The strategy we employ is to replace $P^2$ force correction with a
correction determined by the tree method.
The details of the basic BH tree method as implemented here are presented in
Dubinski (1996) with the only difference being that the force law is
given as the difference between the Newtonian and PM force law given in equation (1).
In brief, the BH tree algorithm  works by
arranging particles in a hierarchy of cubes or cells in an oct-tree structure.
Forces are determined by descending the tree and summing contributions from
interactions with distributions of particles in ``distant'' cells
and nearby particles.
Cell interactions are determined from the quadrupole order
expansion of the potential of the particles within the cube rather than
through direct summation so there are great computational savings.
The trick is to introduce a cell opening criterion based on the ratio of the
size of a cube to the distance of its centre of mass, i.e. an opening
angle.  If the cell is big and nearby, then the opening criterion
is satisfied such that the cell is broken down
into its 8 sub-cells each of which is examined in turn.
If the cell is small and sufficiently distant, the opening criterion
fails and the quadrupole order expansion of the particles in the cell are
used to determine the force and the tree descent ends at this node.
In practice the tree method
requires only $O(N\log N)$ computations per step to determine forces with an
error of 1\% and therefore has a great advantage over direct summation methods.
There are different choices available for the cell opening criterion as
well as strategies for grouping particles to minimize the overhead in tree
walks.  The cell opening criterion and grouping strategy we use here is
the same as that described in Dubinski (1996) with some slight modifications
that we discuss below.

\begin{figure}
\epsfxsize\hsize\epsffile{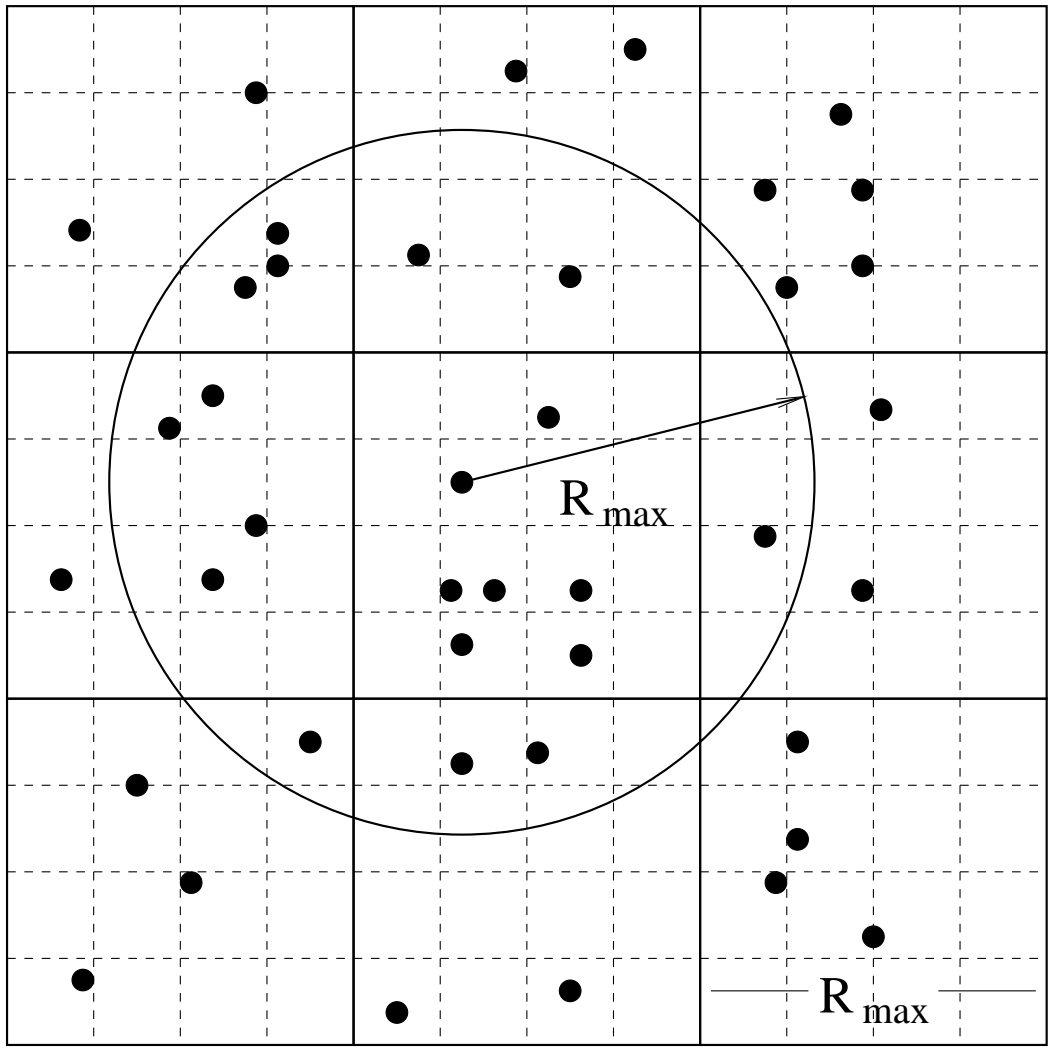}
\caption{
Schematic diagram of the implementation of short range tree forces in 
GOTPM.  In a P$^3$M algorithm, the short range force is determined through
direct summation of forces on particles within a sphere with radius $R_{max}$.
In GOTPM, particles are binned in cubes of width $R_{max}$ and Barne-Hut
trees are constructed from the binned particles with a tree for each cube.
Short range forces are determined by walking through the particle's home tree
and surrounding 26 trees in 3D.  A spherical boundary is not required for
large enough values of $R_{max}$ since the contribution from distance cells
becomes small.  In practice, we need to use $R_{max} = 4 \Delta x$ where
$\Delta x$ is the mesh division.
}
\label{fig-trees}
\end{figure}

The application of the tree method to determine force corrections in a PM
code requires a few general modifications.  The force correction effectively drops to
zero at a few mesh lengths. It is therefore not
necessary to build a global tree for the entire particle distribution in
the simulation cube.   Since the forces are localized, we can build a grid
of trees, with a dimension of a few mesh lengths.  If we define $R_{max}$ as
the radius where the correction forces drops to small fraction of a percent
then we only need to
calculate the contributions from particles within this radius.  In P$^3$M,
one generally sets $R_{max}=2\sim 3$ but we find in practice for the tree
algorithm that $R_{max}=4$ is necessary to assure accuracy.
This scale is set to the length of the cubes enclosing the trees.
Each particle is incorporated into one tree so to determine the force correction 
on a particle we need only descend the
particle's home tree and the surrounding 26 trees (Figure \ref{fig-trees}). 

In principle, P$^2$ force corrections are necessary from
all particles within the volume although particles beyond $R_{max}$
contribute little to the correction.
In practice, particles beyond $R_{max}$ are simply ignored
saving a lot of time.  This is necessary since the
number of particles surrounding the point where 
the potential is calculated
grows rapidly with radius i.e. 
$N \sim r^3$ for a homogeneous
system and $N \sim r$ when clusters with densities of $\rho \sim
r^{-2}$ develop.  In a tree code, however, the number
of interactions does not grow rapidly with radius.  Distant particles are
grouped into larger and larger cells so the number of cell interactions
only grows roughly as $N \propto \log r$ under most circumstances.
It is therefore not necessary to impose a hard spherical boundary
of $R_{max}$ as is done in the $P^3M$ method.
As long as the trees are $R_{max}$ in size then
all effective contributions to the force correction will be accounted for
and the extra forces from the irregular jagged region beyond the spherical
boundary will be small and generally average out.
The force corrections themselves will have an inherent error which depends on the
value of the critical opening angle in the tree algorithm and the value of
$R_{max}$.  
In practice, we show below that as
clustering develops the number of particle cell interactions only grows
slowly from the initial homogeneous state to the strongly
clustered regime.  Computationally, the CPU time per step only grows by a
factor of three from the beginning to the end of a simulation.

Since the trees are localized and compact there is one further refinement that can be
added to improve accuracy and efficiency.  If the number of particles within 
a cube is small, the cost of building and walking the tree for force determinations
can be greater than simply doing a direct forces summation.  We introduce an additional
parameter, $N_{tree}$, that defines the minimum number of particles required for
a tree build.  If the number of particles in a bin is less than $N_{tree}$ then we
do a direct P$^2$ summation for all particles within the radius $R_{max}$. Otherwise,
we use the tree machinery for the force
calculation.  In this mode, a typical run behaves as a P$^3$M code at early times
while at late times trees are used to calculate forces wherever clusters develop.

Since force corrections are local, we see that this algorithm is easily
parallelizable within the context of the PM code we have described.
We outline the implementation of this algorithm below.

\subsection{Slice domain decomposition}

The parallel PM code we describe above distributes the particles
into domain slabs of equal width and assigns them to independent processors
on a parallel computer.  We retain this slab domain decomposition scheme for
force corrections with trees but relax the constraint of equal widths
for the slices to achieve better load balance.  As the system
evolves, particles cross these boundaries and must be moved between
processors by exchanging messages.

The inner loop of the algorithm is as follows:
\begin{enumerate}
\item Particles are redistributed into domain slices of equal width by
exchanging messages for the next PM force evaluation.
\item The PM force is determined using the slice domain decomposition
parallel method described above.
\item Particles are now redistributed into slices of variable width such that
the amount of treecode computation per slice is equal.  The boundaries of
these domains are determined iteratively using the number of tree force
interactions from the last step.  It is a one-dimensional version of the
orthogonal recursive bisection (ORB) domain decomposition method described
in Dubinski (1996) and used in other parallel tree codes like GADGET (Springel
et al. 2001).
At the end of this step, each processor contains
a list of particles which requires the same amount of computation to
determine tree force corrections.
\item We are now ready to build a local grid of trees for the force
correction.  To do this, we need to import ``ghost'' particles from neighbouring
domains extending a distance $R_{max}$ from the current slice boundaries.
This is accomplished by exchanging particles with neighbouring processors.
These ghost particles along with the local particles are then binned into
cubes of dimension $R_{max}$ in the slice.  To save memory, we only import
the positions (and masses if variable masses are present) for the ghost particles.
\item For each particle bin, we then build a BH tree.
\item We now loop through the local particle list determining the force
correction using tree descents from the surrounding 27 trees centred on
each particle.  We remember the number of interactions required for each
particle for load balancing purposes in step 3 above.
\item Force corrections are now added to the PM force to determine the
corrected Newtonian force in a periodic universe.
\item Particles are pushed forward according to the equations of motion and
a leapfrog integration scheme.
\end{enumerate}

\begin{figure}
\epsfxsize=3.0in\epsfbox{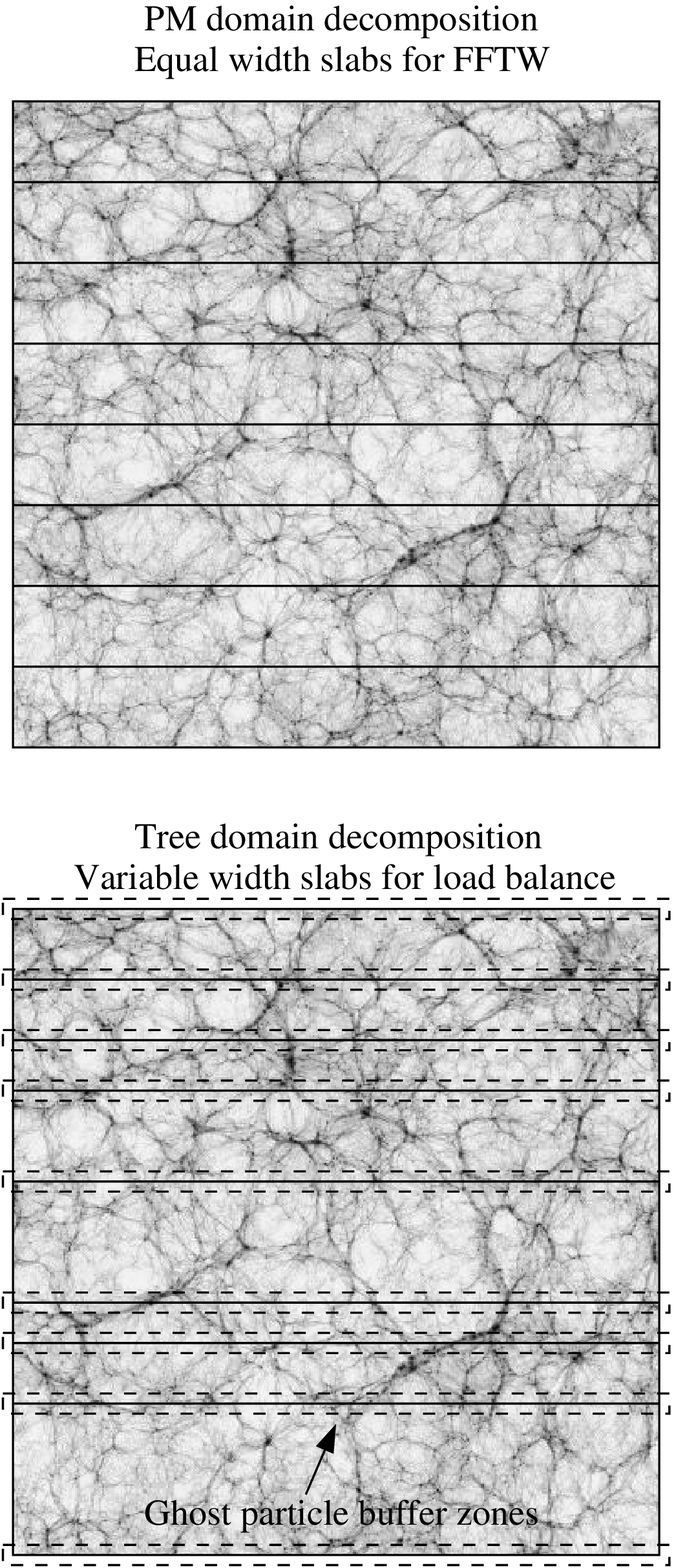}
\caption{
Two modes of slab domain decomposition required to compute PM long
range forces and tree short range force corrections.
The PM method requires particles binned in slabs of equal width
for mass assignment and the use of FFTW to compute the potential.
The tree method requires variable width domains for computational
load balance.  The width of domains is determined by the equalizing
the computational load as measured by the number of particle 
interactions in each slab using a recursive bisection technique.
Ghost particles from buffer zones of width $R_{max}$ on each side
of a slab are imported to build boundary trees.  The periodic boundaries
are also implemented by importing trees from opposite sides of the cube
where appropriate.}
\label{fig-dd}
\end{figure}

Figure \ref{fig-dd} illustrates the domain decomposition scheme needed
for the FFT's and achieve load balance for short range tree forces.

There are a few refinements to this algorithm that improve speed and
minimize memory overhead.  First, step 3 is generally expensive in terms of
communication time since the
domain boundaries must be determined iteratively in $\log_2 N_{proc}$ message
exchanges each possibly requiring the communication of
large numbers of particles.
To speed this step up, the
full ORB decomposition is only done every 
10 steps.
At intermediate
steps the boundaries are held fixed according to the last determination.
Since the system does not change dramatically from step to step,
only a small fraction of the
particles actually cross these boundaries between steps thereby minimizing
communication and preserving load balance.
Second, we save on memory when building the trees
in steps 5 and 6 by only building a small subset of the trees at a time
and discarding them after we are
done with them.
Tree structures inherently require large amounts of data to describe the
relational aspects of the tree through pointers and the cells themselves
require lots of extra information including the quadrupole components as
well as cell dimensional information.
Since all force corrections for a particle are local to the
surrounding 27 trees, we can build the trees in small groups, determine
the correction forces for corresponding particles then discard the trees
before building a new set.  In practice, we build only $8 \times 8 \times n_z$
grid columns of trees and loop through the volume where $n_z$ 
is the number of trees in the vertical direction.  
The slice height $n_z$
varies from step to step because
of the domain decomposition.  For example,
in a typical simulation on a 32 processor machine with a $1024^3$ mesh
with $R_{max}=4.0$, there are on average $256 \times 256 \times 8=512K$ trees
that need to be built in each processor.  We only need to build 512 trees
at a time with this method saving a factor of 1000 in memory for tree
data structures.  The only information 
one needs
to save at each step
is the basic particle data,
i.e. positions, velocities, particle index and
computational load.  Extra memory is required to handle
neighbour particles and memory imbalance induced by the load-balancing
algorithm itself.  For small cosmological volumes with a great deal of
clustering, the memory imbalance leads to roughly a factor of 2 variation
in the number of particles per processor.  In the worst case, we require
about 3 times the memory required for basic particle data (masses, positions and accelerations)
to handle this problem.

\subsection{Memory budget}

The memory budget for a pure n-body simulation can be broken
down into particle memory and mesh memory.  A typical mesh requires
$4\times N_{mesh}^3$ bytes for single precision and twice that for
double precision.   The particles themselves
require memory for positions, velocities, accelerations, masses and softening
lengths, a computational work measurement, a particle index to retain identity
and a pointer for constructing linked lists when binning particles in cubes. 
The total budget is the equivalent of 56 bytes per particle in single precision
or 112 bytes in double precision.  Because of imbalance in particle numbers in
the domain slabs the worst case that occurs with the smallest box sizes
leads to domains with up to 3 times the average number of particles.  
Conservatively, approximately 150 bytes are required per particle in a typical
single precision run.   The ghost particles can potentially eat up significant
amounts of memory when the domain slabs become thin.  This deficiency can
be removed in future versions by using fully 3-D domain decomposition in the
tree correction phase though considerable bandwidth may be required to move
particles back and forth from slabs in the PM phase and cuboids in the tree
phase.

In summary, a typical run with $N_{mesh}=2048^3$ and $N_p=1024^3$ requires $\sim$200 GB
in single precision in the worst case of a small box.  This new algorithm is
considerably more memory efficient than pure cosmological treecodes that require
lots of memory to describe global trees.

\section{Code Verification}

\subsection{Accuracy Tests}

To validate the correctness and accuracy of the code, we calculated the
accelerations from a 2M particle dark matter simulation 
in a $50 h^{-1}$ Mpc box at $z=0$ in a standard cosmology
with $\Omega = 0.3$ and $\Lambda=0.7$.
This resulting particle distribution is strongly clustered and so
represented a good test for the method's accuracy. 
We used several methods to measure the accelerations. 
To start, we used a cosmological parallel treecode based on
the Ewald method
(Dav\'e et al. 1997) to determine accelerations
with small errors by setting the opening 
angle parameter $\theta=0.1$ and computing cell-particles interactions 
to quadrupole order.  We used a Plummer softening length of $5\times 10^{-5}$ times
the box size.  This is a costly calculation but the mean relative error 
in the accelerations $|\delta a/a|$ is estimated
to be $<0.01$\%.  We consider these accelerations
to be exact for the purpose of comparison to other methods.
We then calculated acceleration errors using the Ewald method with a
more standard value of $\theta=0.8$ and the new GOTPM algorithm
with $\theta=0.0$ (making it a $P^3M$ method) and $\theta=1.0$ 
with $N_{tree}=4$ and 32.
The size of the mesh is $256^3$.  In the GOTPM algorithm,
we expect errors to arise both from the treecode approximation as well as
the PM method itself.  The technique of fitting a smooth potential to the
PM force function minimizes but does not eliminate errors induced by close
range anisotropies in the mesh force.

\begin{figure}
\epsfxsize\hsize\epsffile{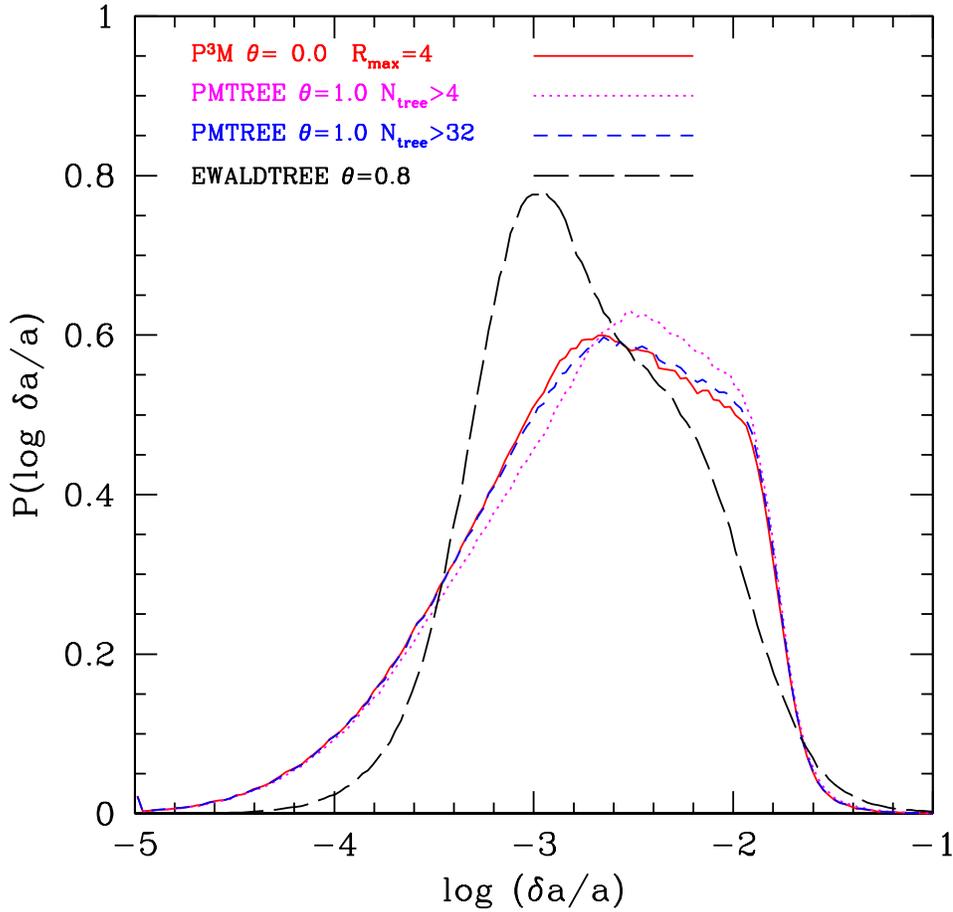}
\caption{
Distribution of relative errors in particle accelerations for 
a clustered 2M particle CDM simulation at z=0 ($\Omega_m=0.3$, $\Lambda=0.7$, 
$L= 50$ h$^{-1}$ Mpc for P$^3$M, GOTPM and a Ewald-method cosmological treecode.
The GOTPM code with $\theta=1$ and $N_{tree}=32$ results in errors similar 
to P$^3$M and comparable to treecodes while using much less computing time.
}
\label{fig-errors}
\end{figure}

Figure \ref{fig-errors} shows the distribution of relative acceleration errors for 
various methods and parameter combinations.
The first result is that the values of accelerations between
the Ewald-tree method and GOTPM algorithm agree with each other on
the percent level.  This result reassures us that the computed
accelerations are physically correct
since the two numerical methods
are very different in their mathematical formulation:  
the first method is based on a combination of 
the Ewald expansion of a periodic
point mass potential and the second method
based on the solution of Poisson's equation using Fourier methods.
We can clearly see the convergence to smaller errors when going from
the P$^3$M to GOTPM method with $\theta=1.0$ and smaller 
values of $N_{tree}$.  The distribution of errors in the P$^3$M represents
the ``best''
one can do to reduce errors in the accelerations using a hybrid
PM method of this form.  
%The errors come from the anisotropy of the grid force at the mesh scale
%and can only be minimized using a smooth correction force.  
The distribution of errors between the P$^3$M method and GOTPM methods 
are almost indistinguishable when $N_{tree} = 32$ while the time taken to compute
these forces using P$^3$M requires about 10 times as much CPU time.
For larger N simulations, the discrepancy in CPU time becomes even greater.
The distribution of errors for a Ewald-method treecode using a typical value
of $\theta=0.8$.  Although, the mean error is slightly less than the GOTPM method there is a
broader tail in the distribution to larger errors precluding the use of a larger value of $\theta$.
The Ewald-method treecode method requires about 3 times as much CPU time to compute accelerations
of comparable accuracy to the GOTPM method in the clustered state.  We note, however, the situation
is very different when the systems are less clustered.   The GOTPM method is several times faster 
and more accurate when the system is less clustered since it operates essentially like a pure 
PM method.   In constrast, the Ewald tree method is only slightly faster in the less clustered case 
and acceleration errors tend to be much larger for a given value of $\theta$.

Using a larger mesh size also increases the
accuracies of the accelerations and as we see below can also requires less
computation so if enough memory exists it can be advantageous to use a finer mesh for
the PM part of the calculation.

Also apparent is a tail of the distribution of errors that remains
as $\theta$ decreases.  The cause of
this tail is due to the anisotropy of the PM force on the mesh
scale which introduces a random error on the determined acceleration for
particles.  This error is inherent to the PM method itself and all
techniques using meshes will have this problem.  We also note that 
the acceleration can be near zero because of the periodic nature
of the force so relative errors are amplified in this case and do
not necessarily indicate low accuracy.
A plot of the acceleration versus the relative error indeed shows that
the largest acceleration errors occur for the smallest accelerations.
For $\theta=1$, the acceleration errors are found to be $0.43\% \pm 0.56\%$,
the range typically tolerated in treecode simulations.  There does not seem
to be a large gain in accuracy in going to smaller $\theta$ because of
mesh-induced force errors.  All methods base on PM forces plus correction will
have these errors so we recommend using $\theta=1$ as a compromise
between speed and accuracy since generally 0.5\% errors are usually tolerated in
collisionless n-body calculations.  We note, however, that the relative
errors are much smaller for particles feeling large accelerations in the
centre of clusters in these simulations so the 0.5\% error is
a conservative estimate.

\subsection{Code Performance}

The GOTPM code has been run on a variety platforms including
a 32 processor IBM SP3 in Seoul, an HP/Compaq GS320 and a Beowulf cluster 
based on Intel Xeon 2.4 GHz chips with gigabit networking both at CITA.
We present results of the code performance and parallel
scaling up to 128 CPUS on the Beowulf cluster though runs with up to
256 CPUS have been done and perform well.  Our reference simulation
computes the evolution of $N=256^3$ particles using a $512^3$ mesh in a 
200 h$^{-1}$ Mpc box in a standard cosmology with 
$\Omega_m=0.3$ and $\Lambda=0.7$.

\begin{figure}
\epsfxsize\hsize\epsffile{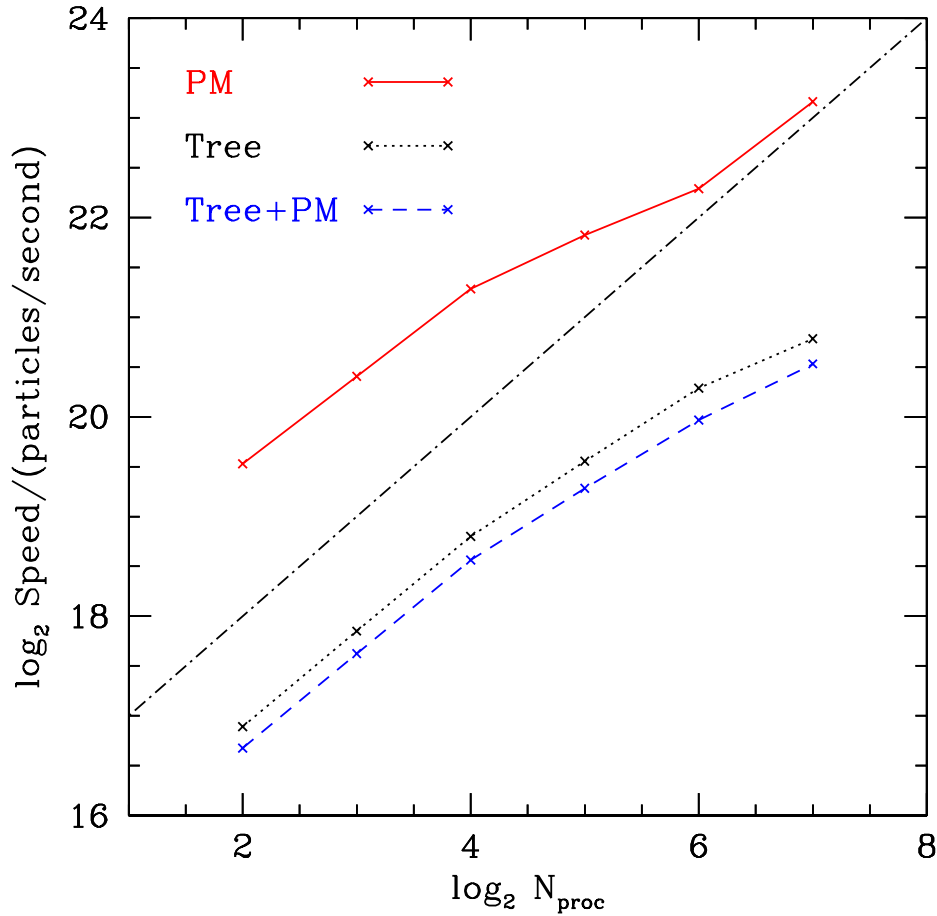}
\caption{
Code speed measured in particles per second versus number of processors
for a $N=256^3$ particle CDM simulation in a $512^3$ mesh with moderate clustering.
The code scales well to 128 processors and the bulk of computing time going
into short range force calculations using the tree method.
}
\label{fig-timing}
\end{figure}

To study scaling of the code speed with processor number, 
we follow the computation for 5 equal time steps from $z=1$ 
when the system is moderately clustered and time the PM and tree phase of the code
separately to determine a speed.   These timing tests are done on a
Beowulf cluster with gigabit networking and dual Xeon 2.4 GHz processors
with code compiled using the Intel C and Fortran compilers.
Figure \ref{fig-timing} shows the code speed versus the number of processors $N_p$ from 4 to 128 processors.
The speed scaling is not perfect since the amount of communication time grows as $N_p$ increases.
As $N_p$ grows, the domain slabs become thinner and the number of
particles crossing boundaries through both particle redistribution for load-balancing
purposes grows significantly as well as the number of ghost particles imported for tree
builds.   Nevertheless, the code scales well to $N_p=128$ requiring only 11.6 second per step
and clocking at about 1.4 million particles/s.  The PM portion of the calculation only 
requires 1.8 seconds compared with 9.8 seconds for the tree correction force
so the code is dominated by determination of the short range forces.

\begin{figure}
\epsfxsize\hsize\epsffile{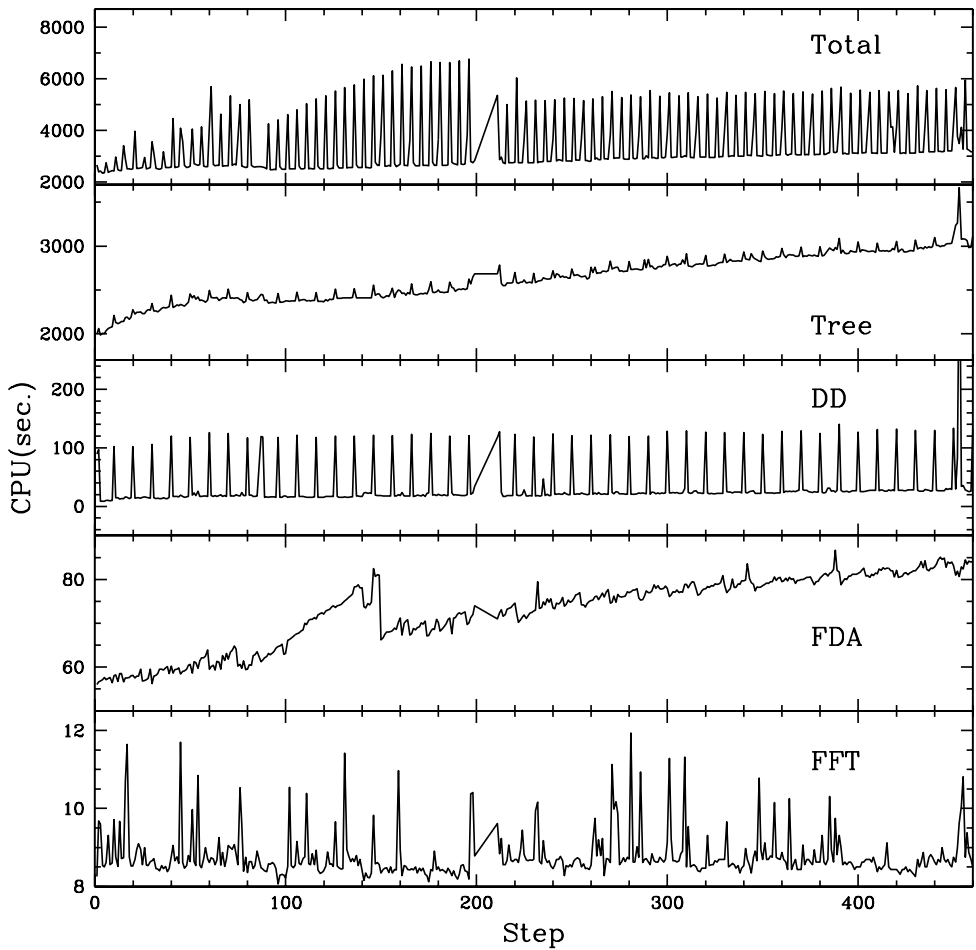}
\caption{CPU time per step in a 1G particle simulation.
Only cpu times of important parts in the code are displayed;
FFT(forward \& backward FFT), FDA(Finite Difference Algorithm),
DD(Domain Decomposition), Tree(Tree force correction).
The jumps in the total timing plot are due to the 
periodic backups of particles every 15 steps and pre-halo-finding 
calculations done every 5 steps.
}
\label{fig-1Gtime}
\end{figure}

\begin{figure}
\epsfxsize\hsize\epsffile{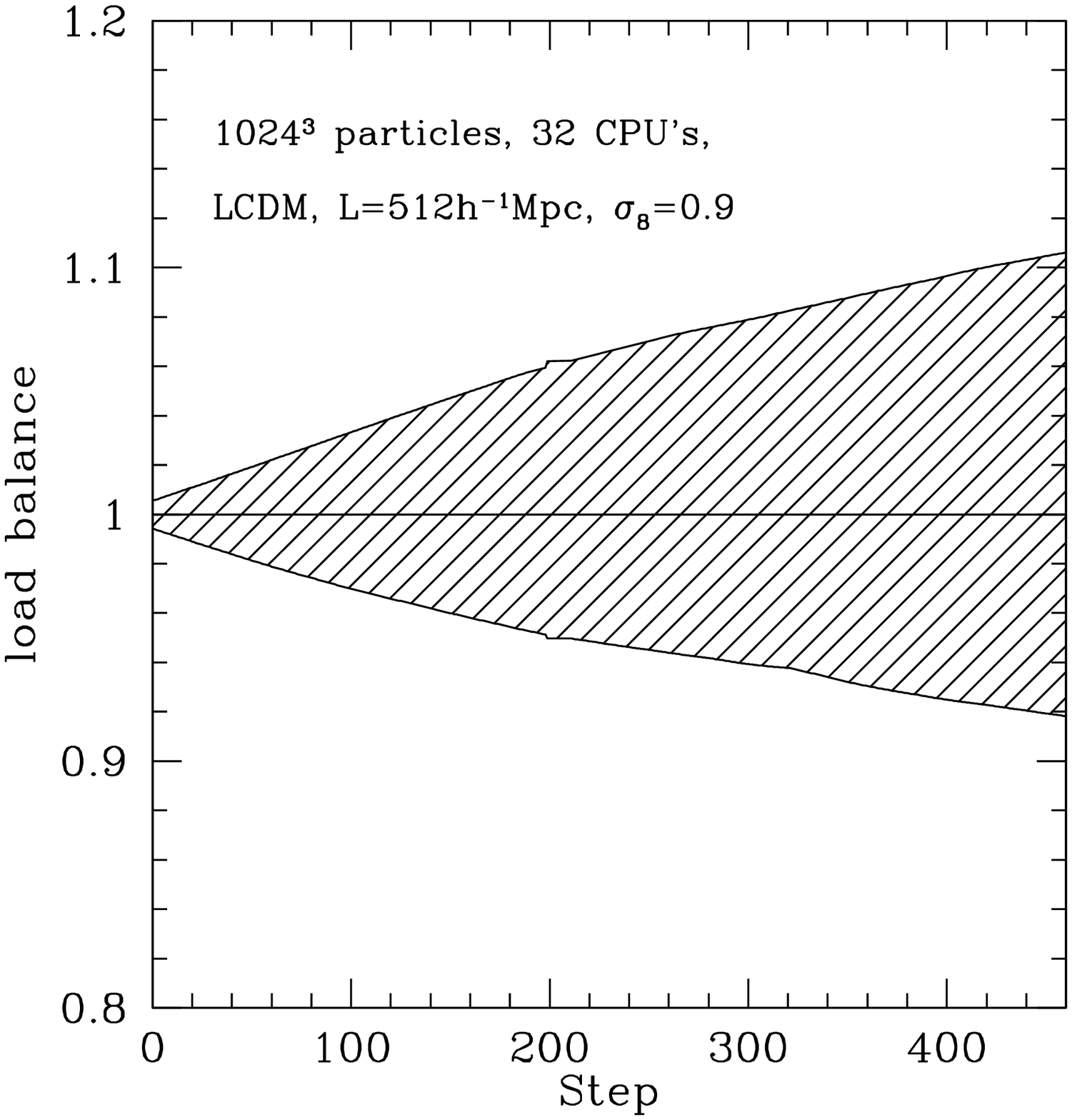}
\caption{Load balance in a 1G particle simulation.
The upper and lower boundaries
of the shaded region indicate
the maximum and minimum (number ratio of particles in a slab 
ratios of domain-particle numbers
to the case of a homogeneous distribution.
}
\label{fig-1Gdd}
\end{figure}

As part of J. Kim's doctoral work,
we simulated $1024^3$ particles in LCDM and SCDM cosmological models
in a cosmological box of size $L=512 h^{-1}$ Mpc.
The simulations were run
on a IBM SP3 (Regatta) system 
at the Korea Institute of Science and Technology Information (KISTI).
Over the course  of two months,
we used one node of 32 CPU's with 208 Gbytes of memory for two
GOTPM simulations.  Memory limited us to a $1024^3$ mesh.
Each model was simulated with 460 time steps in 16 wall-clock days
and the peak memory usage was about 170 Giga bytes.
The simulation details will be presented in Kim's thesis and a 
forthcoming publication.
The CPU timing of the LCDM simulation is displayed 
in Figure \ref{fig-1Gtime}.
The CPU time per step grew from 40 minutes to 55 minutes
largely due to the increase of tree calculation overload 
caused by the clustering of particles.
Periodic spikes in DD and the total CPU time plot are 
due to the particle backup
at every 15 time steps and pre-halo finding calculations every 5 time steps.
The increase of CPU time in the FDA part was partially due to 
the  steady increasing rates of cache miss of particle data in L2 cache.
As the system evolves, the memory locality of the particle data 
differs significantly from the spatial positions of the particles.
In Figure \ref{fig-1Gdd}, we show the inhomogeneity of particle
distributions of PM domain slabs in the LCDM simulation.
We expect the density fluctuation$\Delta(r)$ at length scale $r$
is related to the power index $n$ as (Longair 1998)
\begin{equation}
\Delta(r) = \left( { r \over {8h^{-1}{\rm Mpc}}}\right)^{-{n+3 \over 2}},
\end{equation}
where $r$ is defined to be the radius of a sphere having equal volume of 
the local domain slab.
In the case of $n=-1, -2$, $\Delta(M)$ of the simulation domain slabs 
is $0.16, 0.03$, respectively.
Hence the static domain decomposition of the PM part
does not matter seriously to memory budget overload for cosmological volume.

%As part of a ongoing project to study clustering and galaxy evolution in a small
%volume, we ran a cosmological simulation in a $65h^{-1}$ Mpc box using
%the current cosmological parameters $(\Omega,\Lambda,h,\sigma_8)
%= (0.3,0.7,0.65,1.0)$.  The simulation used $N=512^3$ particles
%with $N_mesh=1024$ for the PM calculations
%and $\theta=1.0$ for the treecode calculations.
%The simulation was run for 2800 timesteps from $z=71$ to $z=0$.
%We timed various portions of the code to get a sense of load balance,
%communication overhead and overall performance.   This run was carried out
%on CITA's 32-processor COMPAQ GS320.
%
%Figure (?) shows the code speed as a function of timestep
%measured by wallclock timing of individual timesteps.  Initially, the
%system is nearly homogeneous.

%If the volume of each PM domain slab is greater than the sphere of radius
%$8h^{-1}{\rm Mpc}$, then the fluctuation of particle number of slabs
%is less than the bias factor in $1\sigma$ limit.

\section{Mass Functions}

\begin{figure}
\epsfxsize\hsize\epsffile{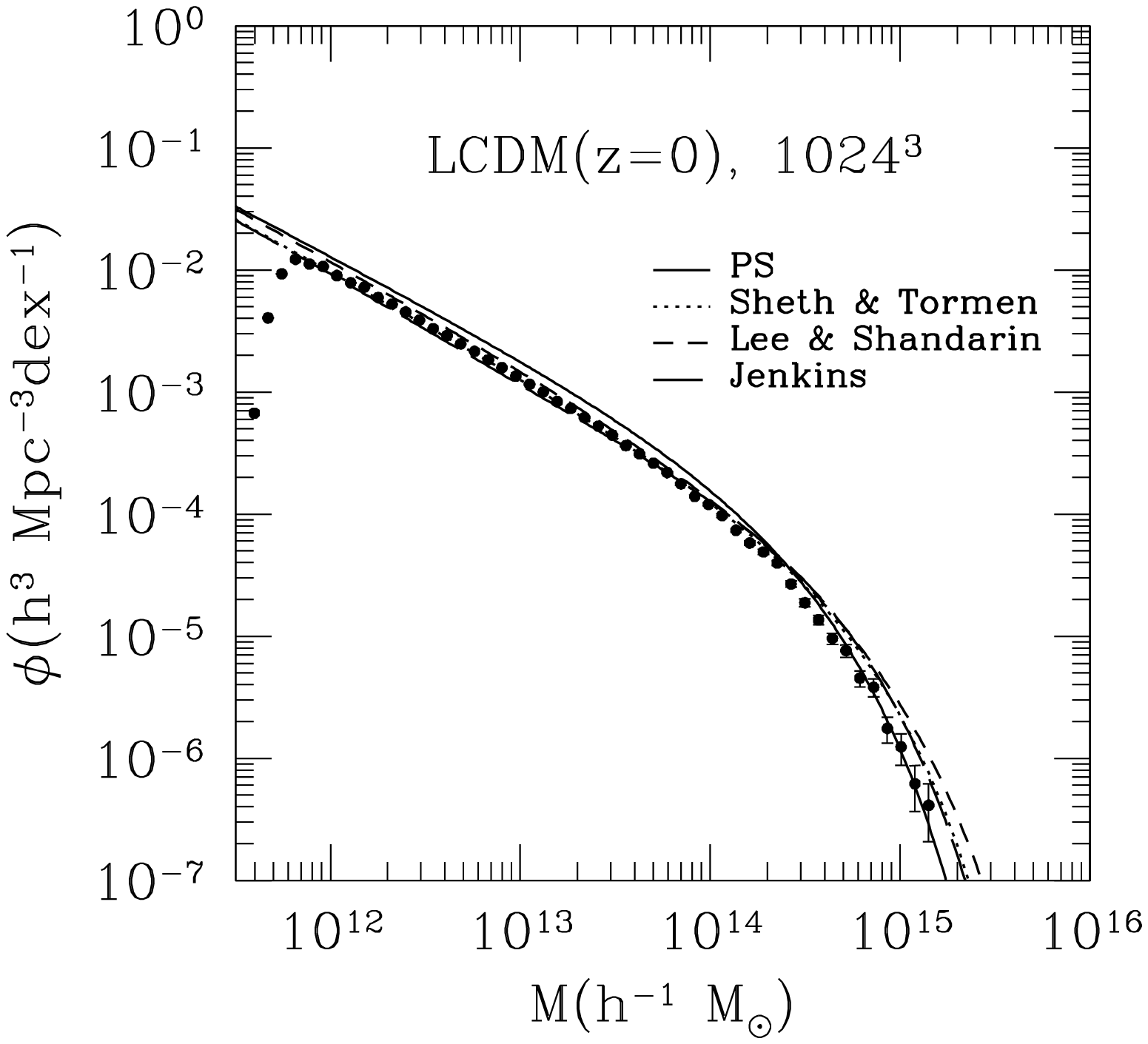}
\caption{ Mass functions ($z=0$)
obtained from a 1G particle simulation of a LCDM model.
We overlap halo mass functions({\sl filled circles})
with a Press-Schecter({\sl a solid line}),
Sheth-Tormen({\sl a dotted line}),
Lee-Shandarin({\sl a dashed line}), and
Jenkins({\sl a long-dashed line}) mass functions.
}

\label{fig-mflcdm0}
\end{figure}

Another useful tool for the verification of the code is
the comparison of simulated halo mass functions 
with other analytic halo mass functions.
There have been several fitting functions proposed by many authors for 
halo mass functions.
Press and Schecter(1974,hereafter PS), Lee and Shandarin(1998,hereafter LS)
Sheth and Tormen(1999,hereafter ST) and Jenkins et al.(2000) have presented
analytic and fitting mass functions.
On the basis of spherical collapses and Gaussian fluctuations
in the initial density field, Press and Schecter derived
the mass function as,
\begin{eqnarray}
\begin{array}{ll}
{\rm d} N(M) = \sqrt{2\over\pi}{\rho_0 \over M}{\delta_c\over\sigma}
\exp\left(-{\delta_c^2 \over 2\sigma^2}\right),
\end{array}
\end{eqnarray}
where $\delta_c$ is defined $1.686$ and $\sigma(M)$ could be
calculated from the initial power spectrum $P(k)$.
Other analytic halo findings are rather empirical functions to fit to
their simulated halo mass functions. 
We display the halo mass function for the 1G particle LCDM simulation 
with other analytic halo mass functions in Figure \ref{fig-mflcdm0}.
In a forthcoming paper, we will describe a halo-finding method
used to extract virialized halos from the simulation data during run time.
The simulated halo mass function of mass 
above $2\times10^{14}h^{-1}{\rm M_\odot}$ is well fitted by the PS predictions
but underestimates halo numbers by about 30\% below this mass scale.
Even with the shortcomings at the high-mass end,
our halo mass functions show global consistency with ST and Jenkins predictions.

%\begin{figure}
%\epsfxsize\hsize\epsffile{figs/fig-mfscdm0.eps}
%\caption{ Mass funcion($z=0$)
%obtained from a 1G particle simulation of a SCDM model.
%}
%\label{fig-mfscdm0}
%\end{figure}

We also note that the PM version of the code has already been used to do a series of
$512^3$ particle simulations to model gravitational lensing convergence and shear 
for comparison to lensing surveys (Pen et al. 2003).

\section{Conclusions}

GOTPM is a new parallel hybrid Tree/PM cosmological N-body method that is fast, accurate, 
memory efficient and scaleable to
hundreds of processors on parallel machines.  We have verified the correctness and accuracy of the forces through
comparison with other methods and produced a mass function consistent with other results in
the literature.  This new code is well-suited for the next generation of large-scale
cosmological simulations on massively parallel machines.  

At present the code has been used to compute $N=1G$ particle simulations but to meet the challenges
of doing even larger calculations on future supercomputers with thousands of processors some modifications
will be necessary.  The current version uses a single leapfrog timestep.   The incorporation
of hierarchical multiple timestepping should be straightforward.  Since the rapidly changing force field
in dense regions will be mainly calculated as a tree correction, the PM part of the potential can be held
fixed or extrapolated over the largest system time step.  Since all near range forces are localized to 
particles in the algorithm only trees in the densest regions need to be updated at smaller fine-grained
timesteps while remaining trees can be repaired using methods described by Springel et al. (2001).
Furthermore, the communication needs will be small at the fine-grained timesteps since forces are all determined
locally and few particles will cross domain boundaries.  Work is proceeding to incorporate multiple
timestepping.

One major weakness of the algorithm from the perspective of parallelization and memory needs is 
the use of slab domain decomposition for the tree correction phase.  As the number of CPU's increase
slabs become thinner especially when the system becomes more inhomogeneous.  This leads to dramatic
increase in both communication to move ghost particles as well as memory usage since the imported ghost
particles can actually dominate local memory.  For $N=1G$ simulations with up to 256 processors the algorithm works
well but slab domain decomposition will break down on future machines with thousands of processors. 
One solution to this weakness, is the use of 3-D cuboid domain decomposition instead of slabs for the
tree phase.  This will solve the ghost particle memory and communication problem.  The PM phase will still
require slabs since this is inherent to FFTW's MPI implementation.  Either particle data will have to be moved
back and forth between cuboids and slabs every step or the mesh points themselves will have to
be communicated.  These various strategies will be examined in coming versions and will prepare the way
for $N=8-64$G particle simulations.

Another weakness of the algorithm is the scatter in the PM mesh force anisotropy.
We can minimize this problem by introducing a spherically symmetric mass allocation
kernel.  The extra cost of this scheme will still be small compared to the
tree phase of the algorithm.  The advantage of this method is that we can potentially
reduce the size of $R_{max}$ and save on computation during the tree force corrections.

The code is also well suited for incorporation of smoothed particle hydrodynamics (SPH).  The grid
of localized oct-trees can be exploited immediately to form neighbour lists for SPH estimation of density and
pressure fields.  With the multiple timestepping scheme described above, GOTPM with SPH is a natural fit.

The code is currently available by request but will be released publicly in the near future.

\section*{Acknowledgments}
JD thanks NSERC and the Canadian Foundation for Innovation
and acknowledges the CITA pre-doc programme that allowed JK's visit.
JK and CP thank the Brain Korea 21 program of the Korean Government and
the Basic Research Program of the Korea Science \& Engineering Foundation
(grant no. 1999-2-113-001-5). JK and CP also acknowledge the support 
from the KISTI (Korea Institute of Science and Technology Information) 
under `Grand Challenge Support Program' with Dr. S. M. Lee as the technical supporter.
The use of the computing system of the Supercomputing Center is also greatly appreciated.

\end{document}